\documentclass[conference]{IEEEtran}
\IEEEoverridecommandlockouts
\usepackage{cite}
\usepackage{amsmath,amssymb,amsfonts}
\usepackage{algorithmic}
\usepackage{graphicx}
\usepackage{textcomp}
\usepackage{xcolor}
\usepackage{setspace}
\def\BibTeX{{\rm B\kern-.05em{\sc i\kern-.025em b}\kern-.08em
		T\kern-.1667em\lower.7ex\hbox{E}\kern-.125emX}}
\usepackage{subfigure}

\newtheorem{my_lemma}{Lemma}

\newtheorem{my_proposition}{Proposition}

\usepackage[margin=0.7in]{geometry}
\title{Simplified Performance Analysis of OWC System Over  Atmospheric Turbulence with Pointing Error}
\author{
	\IEEEauthorblockN{Kartik Wardhan and S. M. Zafaruddin}\\
	\IEEEauthorblockA{ Deptt. of Electrical and Electronics Engineering, 
		BITS Pilani, Pilani-333031, Rajasthan, India.\\ Email: \{f20170301, syed.zafaruddin\}@pilani.bits-pilani.ac.in}
	
	\thanks{  }
}

\begin{document}
	\maketitle
	\begin{abstract}
Optical wireless communication (OWC)  is highly vulnerable  to the  atmospheric turbulence and pointing error.  Performance analysis  of the OWC system under the combined  channel effects of pointing errors and atmospheric turbulence is desirable for its efficient deployment.  The widely used Gamma-Gamma statistical model for atmospheric  turbulence, which consists of Bessel function,  generally leads to complicated analytical expressions. In this paper, we consider the three-parameter exponentiated Weibull model for the atmospheric turbulence to  analyze the ergodic rate and average signal-to-noise ratio (SNR) performance of a single-link OWC system. We derive simplified analytical expressions on the  performance under the combined effect of  atmospheric turbulence and pointing errors in terms of system parameters. We also derive approximate expressions on the performance  under the atmospheric turbulence by  considering negligible pointing error. In order to evaluate the performance at high SNR, we also develop asymptotic bounds on the average SNR and ergodic rate for the considered system.  We demonstrate the tightness of derived expressions through numerical and simulation analysis along with a comparison to the performance obtained using the  Gamma-Gamma model. 
	 
	\end{abstract}

	\section{ Introduction }
	
	Optical wireless communication (OWC) is a promising technology with applications in many fields such as broadband data transmission, last-mile access, and high-speed wireless backhaul \cite{Khalighi2014,Kedar}. The OWC enjoys enormous bandwidth in the license-free spectrum thereby providing high data rate transmissions under the line-of-sight (LOS) channel conditions. However,  OWC links are highly vulnerable  to the  atmospheric turbulence caused by the scintillation effect of light propagation over unguided medium \cite{Li,Zhu2002}. The atmospheric turbulence deteriorates the link performance by inducing fluctuations in the intensity and the phase of received optical beams. In addition to this, pointing errors  can significantly degrade the performance of the OWC system. The pointing error is the  misalignment between the transmitter and receiver caused by the thermal expansions, dynamic wind loads and weak earthquakes resulting  in the building sway and  mechanical vibration of the transmitter beam \cite{Farid2007,Ruiz2016}. Performance analysis of OWC system under the combined  channel effect of pointing error and atmospheric turbulence is desirable for its efficient deployment.

	There has been  extensive research to analyze the OWC performance under the atmospheric turbulence and pointing errors by deriving analytical bounds on various metrics such as outage probability, bit-error-rate (BER), average signal-to-noise ratio (SNR), and ergodic capacity \cite{Zhu2002, Arnon2003_Opt, Farid2007, Nistazakis2009,Kaur2013}.  These analyses use statistical fading models for  the intensity fluctuations  and pointing errors. Assuming independent identical distributed Gaussian  for the elevation and the horizontal displacement and considering the effect of beam width and detector size, a pointing error model, was proposed in  \cite{Farid2007}.  This model is widely used in the literature. However, there are  quite a few  statistical models for the atmospheric turbulence, for example, log normal \cite{Zhu2002}, Gamma-Gamma (GG) \cite{Andrews20015}, and Mal\'aga \cite{malaga2011}. The lognormal model is restricted to  weak turbulence conditions for a point receiver \cite{Vetelino2007}.   The GG model has gained  wide acceptance for moderate-to-strong turbulence regime \cite{Ammar2001} whereas the  Mal\'aga is a more generalized model considering all irradiance conditions in homogeneous and isotropic turbulence \cite{Malaga2012}. Performance bounds under these channel models mostly consist of complicated mathematical functions and generally  do not provide insights on the system behavior. Further, under aperture averaging conditions, it has been observed that these statistical models often do not provide a good fit to simulation data in the moderate-to-strong turbulence regime  \cite{Lyke2009}.  It is noted that aperture averaging is an effective technique 	to mitigate atmospheric turbulence with a large 	collecting aperture detector in the OWC link.
		
	Recently, Barrios and Dios \cite{Barrios2012}  proposed  the exponentiated Weibull (EW) distribution model for the atmospheric turbulence. This model provides a good fit between simulation and experimental data under moderate-to-strong turbulence for aperture averaging conditions, as well as for point-like apertures. A distinguishing  feature of the EW model is its simple closed-form	expression of the probability distribution function (PDF). This has sparked research interest to analyze  the OWC performance and derive closed-form expressions on the outage and BER in \cite{Yi2012, Wang2014, Sharma2015, Boluda-Ruiz2017,Agarwal2019} and ergodic rate in  \cite {Cheng2014, Wang2015,Wang2016,agarwal2018}
	 which was not readily feasible using other statistical models.  However, derived analytical expressions for ergodic rate is generally represented in
	  Meijer G-function. Further, average SNR performance of the OWC performance under turbulence channel even  without pointing error is not available in the literature. Simplified performance bounds on the ergodic rate and average SNR is desirable for real-time tuning of system parameters for  efficient deployment of OWC systems.

	In this paper, we analyze the average SNR and ergodic rate performance of a single-link OWC system. First, we derive approximate expressions on the performance  under the atmospheric turbulence by  considering negligible pointing error.  Then, we derive simplified closed-form expressions on the ergodic rate and average SNR under the combined  effect atmospheric turbulence and pointing errors in terms of system parameters.  To further simplify the analysis, we develop asymptotic bounds on the average SNR and ergodic rate useful in the high SNR regime. The derived expressions are simple and do not contain complicated mathematical functions. We perform extensive numerical and simulation analysis to demonstrate the accuracy of derived analytical expressions and compare the performance obtained using the complicated Gamma-Gamma channel  model.

	\section{System Model}
	We consider a single-link OWC system that  employs intensity modulation direct detection (IM/DD) technique for signal transmission. The information is transmitted by the variations in the intensity of the emitted light which is detected at the receiver by a photo-detector.	The signal received at the detector of an OWC system can be  represented as 
	\begin{eqnarray}
	y = hRx + w
	\label{eq:received_signal}
	\end{eqnarray}
	where $y$ is the received signal, $R$ is detector responsivity, $x$ is the transmitted signal, $h$ is the random channel attenuation, and $w$ is the additive white Gaussian noise (AWGN) with zero mean and variance $\sigma_w^2$.  Assuming that the OWC channel is flat fading,  an expression for SNR $\gamma$ is:
	\begin{eqnarray}
	\gamma= \frac{2 P_{\mathrm{opt}}^{2} {R^2}{h^2}}{\sigma_w^2}=\gamma_0 h^{2}
	\label{eq:gamma_0}
	\end{eqnarray}
	where $ \gamma_0= \frac{2P^2_tR^2}{\sigma^2_w}$ and $P_{\mathrm{opt}}$ is the average transmitted optical power such that $x\in \{0,2P_t \}$.
	
	 The channel parameters $h = L h_{a} h_{p}$ consists of three main factors:   path loss $(L)$, atmospheric turbulence $(h_a)$, and  pointing errors $(h_p)$.
 	The atmospheric path loss $L$ is a deterministic quantity defined by the exponential Beer-Lambert law as $L = e^{-\phi d_{}}$, where $d_{}$ is the link distance (in \mbox{m}) and $\phi$ is the atmospheric attenuation factor which depends  on the wavelength and visibility range \cite{Issac2001}. For a low viability range i.e., in foggy conditions, the path loss becomes a random quantity \cite{Esmail2017_Access2, Rahman2020,rahman2020cl}. The factor $h_a$ is the random  atmospheric turbulence channel state  with PDF \cite{Barrios2012}:
	\begin{equation}
	\begin{aligned} f_{h_a}\left(h_a\right)=& \frac{\alpha \beta}{\eta}\left(\frac{h_a}{\eta}\right)^{\beta-1} \exp \left[-\left(\frac{h_a}{\eta}\right)^{\beta}\right] \\ & \times\left\{1-\exp \left[-\left(\frac{h_a}{\eta}\right)^{\beta}\right]\right\}^{\alpha-1}, h_a \geq 0 
	\end{aligned}
	\label{eq:pdf_ha}
	\end{equation}
	where $\beta > 0$ is the shape parameter of the scintillation index (SI), $\eta > 0$ is a scale parameter of the mean value of the irradiance and $\alpha > 0$ is an extra shape parameter that is strongly dependent on the receiver aperture size. The specific values of the parameters $\alpha$, $\beta$ and $\eta$ as well as some expressions for evaluating these parameters is given in \cite{Barrios2012}. 
	
	The PDF of pointing errors fading $h_{p}$ is \cite{Farid2007}:
	\begin{equation}
	\begin{aligned}
	f_{h_p}(h_p) &= \frac{\rho^2}{A_{0}^{\rho^2}}h_{p}^{\rho^{2}-1}, \quad  0 \leq h_p \leq A_0,
	\end{aligned}
	\label{eq:pdf_hp}
	\end{equation}
	where $A_0=\mbox{erf}(\upsilon)^2$ with $\upsilon=\sqrt{\pi/2}\ a/\omega_z$ and $\omega_z$ is the beam width,	and $\rho = {\frac{\omega_{z_{\rm eq}}}{2 \sigma_{s}}}$ with  $\omega_{z_{\rm eq}}$ as the equivalent beam width at the receiver and $\sigma_{s}$ as the variance of pointing error displacement characterized by the horizontal sway and elevation \cite{Farid2007}.

	\section{Performance Analysis}
	In this section, we analyze the average SNR and ergodic rate performance of the OWC system under the atmospheric channel considering both the cases with and without pointing error. 
	
		The average SNR 	$\bar{\gamma}$ and ergodic rate $\bar{C}$ of the OWC system is defined as:
	\begin{eqnarray}
		\label{siso_snr}
	\bar{\gamma}&=\int\limits_{0}^{\infty} \gamma f_\gamma(\gamma) d\gamma\\
	\bar{C}&=\int\limits_{0}^{\infty}  \log_2 (1+\gamma) f_\gamma(\gamma) d\gamma.
	\label{siso_rate}
	\end{eqnarray}
	where $f_{\gamma}\left(\gamma\right)$ denotes the PDF of SNR $\gamma$. In what follows, we derive simplified expressions on  	\eqref{siso_snr} and 	\eqref{siso_rate} using the distribution function $f(\gamma)$.
\subsection{Atmospheric Turbulence Channel}
	First, we consider the impact of atmospheric turbulence channel on the average SNR and ergodic rate performance on the OWC system. Assuming $h_p=1$ and  substituting $h_a =\sqrt{\frac{\gamma}{{\gamma_0} L^2 }}$ in \eqref{eq:pdf_ha}, we get the PDF of SNR for the OWC system under the combined effect of path loss and atmospheric turbulence:
	\begin{eqnarray}
		{f_{\gamma}(\gamma)} =&\frac{\alpha \beta}{2 \eta \sqrt{\gamma {\gamma_{0} L^2}}}  \Big(\frac{\sqrt{\gamma}}{ \eta \sqrt{\gamma_{0} L^2}}\Big)^ {\beta -1}\exp\Big[-{\Big(\frac{\sqrt{\gamma}}{ \eta \sqrt{\gamma_{0} L^2}}\Big)^ {\beta}}\Big]\nonumber \\
	&\Big[1- \exp\Big[-{\Big(\frac{{\sqrt{\gamma}}} { \eta \sqrt{\gamma_{0} L^2}}\Big)^ {\beta}}\Big]   \Big] ^{\alpha -1}                                	
	\label{eq:pdfgamma}
	\end{eqnarray}
	
Note that the direct application of \eqref{eq:pdfgamma} in \eqref{siso_snr} and \eqref{siso_rate} is intractable to derive close form expressions for the average SNR and ergodic capacity.
	\begin{my_lemma}
		\label{snr_all_k1}
		If  $\alpha$, $\beta$, and $\eta$ are the parameters of exponentiated Weibull turbulence channel and $L$ is the path loss of the OWC link, then  approximate expressions for the average SNR and ergodic rate are given as 
		\begin{eqnarray}
		&	\bar{\gamma}\approx \frac{\alpha(\beta+1)}{\beta}  \eta^{2-\beta}\big(\frac{1}{\sqrt{\gamma_0L^2}}\big)^{\beta-2} \Gamma (\beta+1)\nonumber \\
		& \Big[\big(\frac{\beta \frac{1}{\sqrt{\gamma_0 L^2}}}{\eta}\big)^{-\beta}-\frac{(\alpha-1)^2 \beta^4 \big(\frac{\frac{1}{\sqrt{\gamma_0 L^2}} ((\alpha-1) \beta^2+1)}{(\alpha-1) \beta \eta}\big)^{-\beta}}{((\alpha-1) \beta^2+1)^2}\Big]
			\label{eq:avgSNR_approx_formula}
		\end{eqnarray} 
			\begin{eqnarray}
&	\bar{C} \approx   \frac{\alpha \beta  \zeta}{\log 4}  {\eta^{-\beta}}  \big(\frac{1}{{\sqrt{\gamma_0 L^2}}}\big)^{\beta} \Big[2 \Gamma \Big(\beta+\frac{2}{\zeta}\Big) \nonumber \\		& \times  \Big(\big(\frac{\beta \frac{1}{\sqrt{\gamma_0 L^2}}}{\eta}\big)^{-\frac{\beta \zeta+2}{\zeta}}-\Big(\frac{\frac{1}{\sqrt{\gamma_0 L^2}} (\alpha \beta^2+2)}{\alpha \beta \eta}\Big)^{-\beta-\frac{2}{\zeta}}\Big)\nonumber \\
		&+ 2 \Gamma (\beta) \Big( \big(\frac{\frac{1}{\sqrt{\gamma_0 L^2}} (\alpha \beta^2+2)}{\alpha \beta \eta}\big)^{-\beta}- \big(\frac{\beta \frac{1}{\sqrt{\gamma_0 L^2}}}{\eta}\big)^{-\beta}\Big)\Big]
				\label{eq:erg_cap_approx_formula}
		\end{eqnarray}
		
		where  $\zeta$ is a positive integer.	
	\end{my_lemma}
	
	\begin{IEEEproof}	We use an approximation $(1-\exp[-x^a])^b\approx 1- \exp[-x/ab]$  (which is verified extensively through simulations for parameters under consideration) in  \eqref{eq:pdfgamma}  to get
	\begin{eqnarray}
		{f_{\gamma}(\gamma)} \approx& \frac{\alpha \beta}{2 \eta \sqrt{\gamma {\gamma_{0} L^2}}}  \Big(\frac{\sqrt{\gamma}}{ \eta \sqrt{\gamma_{0} L^2}}\Big)^ {\beta -1}\exp\Big[-{\Big(\frac{\sqrt{\gamma}}{ \eta \sqrt{\gamma_{0} L^2}}\Big)^ {\beta}}\Big]\nonumber \\
	&\left[1-\exp\Big[{-\frac{\sqrt{\frac{{\gamma}}{L^2 \gamma_{\text{o}}}}  }{(\alpha-1) {\beta} {\eta}}}\Big]\right] 
		\label{eq:pdfgamma_ap_snr}
	\end{eqnarray}
Using \eqref{eq:pdfgamma_ap_snr} in  (\ref{siso_snr}), we get \eqref{eq:avgSNR_approx_formula}. 
		Similarly, to get an expression for the ergodic capacity, we use the inequality $	\log(1+\gamma)\geq \log\gamma\leq \zeta ({\gamma^{\frac{1}{\zeta}}} - 1) $, where $\zeta$ is a positive integer [\cite{Abramowitz1972book}, 4.1.37]	in  (\ref{siso_rate}), and apply standard procedures to get (\ref{eq:erg_cap_approx_formula}).
	\end{IEEEproof} 
	
In order to derive asymptotic expression,  we consider the asymptotic PDF of the EW turbulence \cite{Boluda-Ruiz2017}:
	\begin{eqnarray}
		f_{h_a}\left(h_a\right)= \frac{\alpha \beta}{\eta^{\alpha \beta}} h_{a}^{\alpha \beta - 1},  0 \leq h_{a} \leq \eta
		\label{eq:pdf_ha_asymp}
	\end{eqnarray}
		Substituting $h_{a} = \sqrt{\frac{\gamma}{L^2 \gamma_0}} $, we get an asymptotic PDF of the SNR:
		\begin{eqnarray}
		f_{\gamma}(\gamma)= \frac{\alpha \beta}{2   \eta^{\alpha \beta} L^{\alpha \beta}   \gamma_{o}^{\frac{\alpha \beta}{2}}} \gamma^{\frac{\alpha \beta - 2}{2}}, 0 \leq \gamma \leq \eta^{2} \gamma_{o}
		\label{eq:pdf_SNR_atm_asymp}
	\end{eqnarray}

	\begin{my_proposition}
		\label{snr_erg_cap_atm_asymp}
		If  $\alpha$, $\beta$, and $\eta$ are the parameters of exponentiated Weibull turbulence channel and $L$ is the path loss of the OWC link, then  asymptotic expressions for the average SNR and ergodic rate are given as 
				\begin{eqnarray}
				\label{eq:avgSNR_symp_formula}
		&	\bar{\gamma} = \frac{\alpha \beta \eta^2}{2 + \alpha \beta} \gamma_{o}\\
						&	\bar{C} = \frac{2}{\alpha \beta \log(4)} \Big( -2 + \alpha \beta \log(\eta^2 \gamma_{o}) \Big)
		\label{eq:erg_cap_asymp_formula}
		\end{eqnarray}
		
	\end{my_proposition}
	
	\begin{IEEEproof}
It is straightforward to prove by substituting 	\eqref{eq:pdf_SNR_atm_asymp} in 	\eqref{siso_snr} and 	\eqref{siso_rate}.
	\end{IEEEproof}

\begin{figure*}[t]
	\centering
	\subfigure[SNR at different $d$ with $C_n^2= 8 \times 10^{-14}$.]{\includegraphics[scale=0.195]{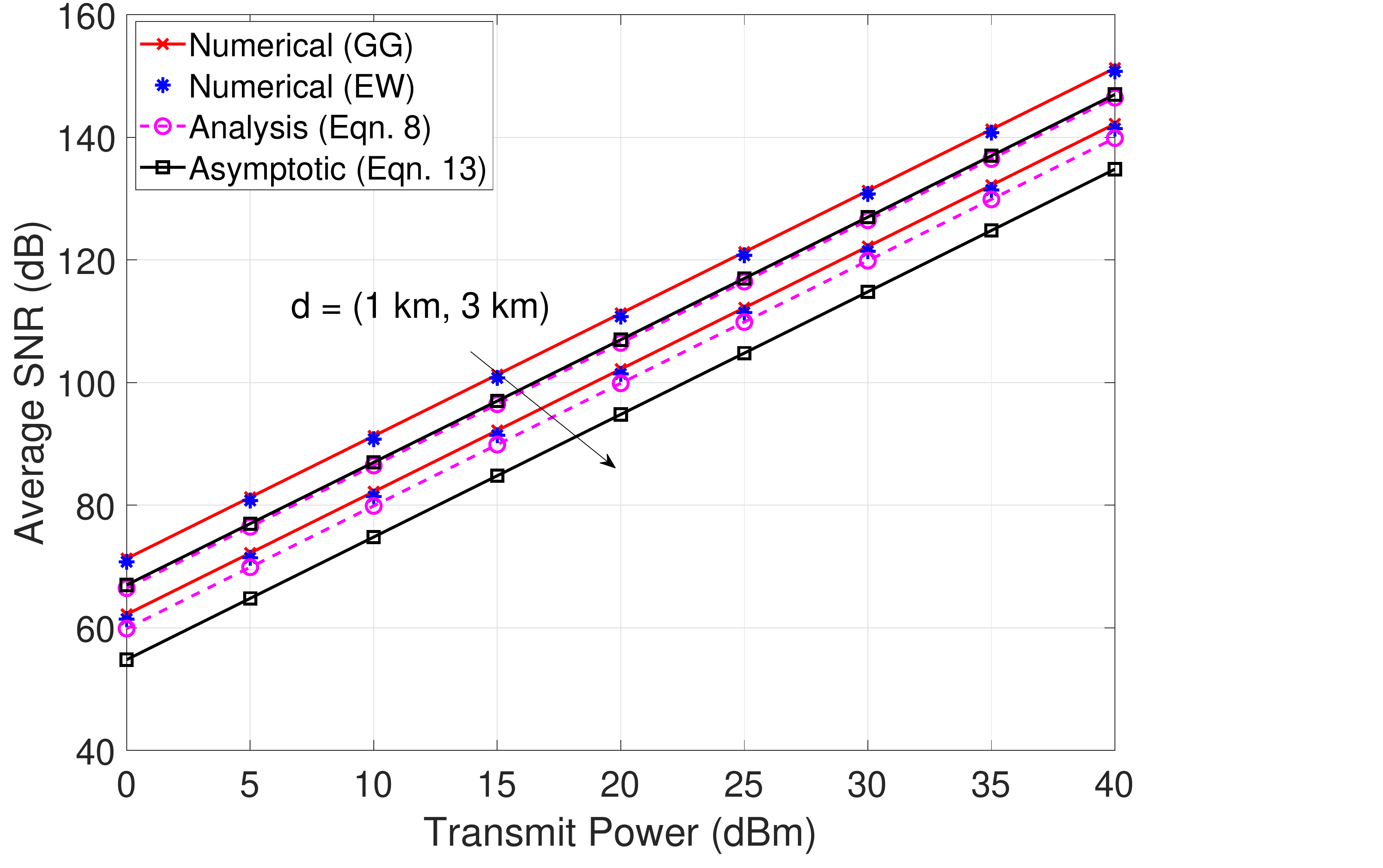}}\hspace{-2em}%
	\subfigure[SNR at different $C_n^2$ with $d=2$ \mbox{km}.]{\includegraphics[scale=0.19]{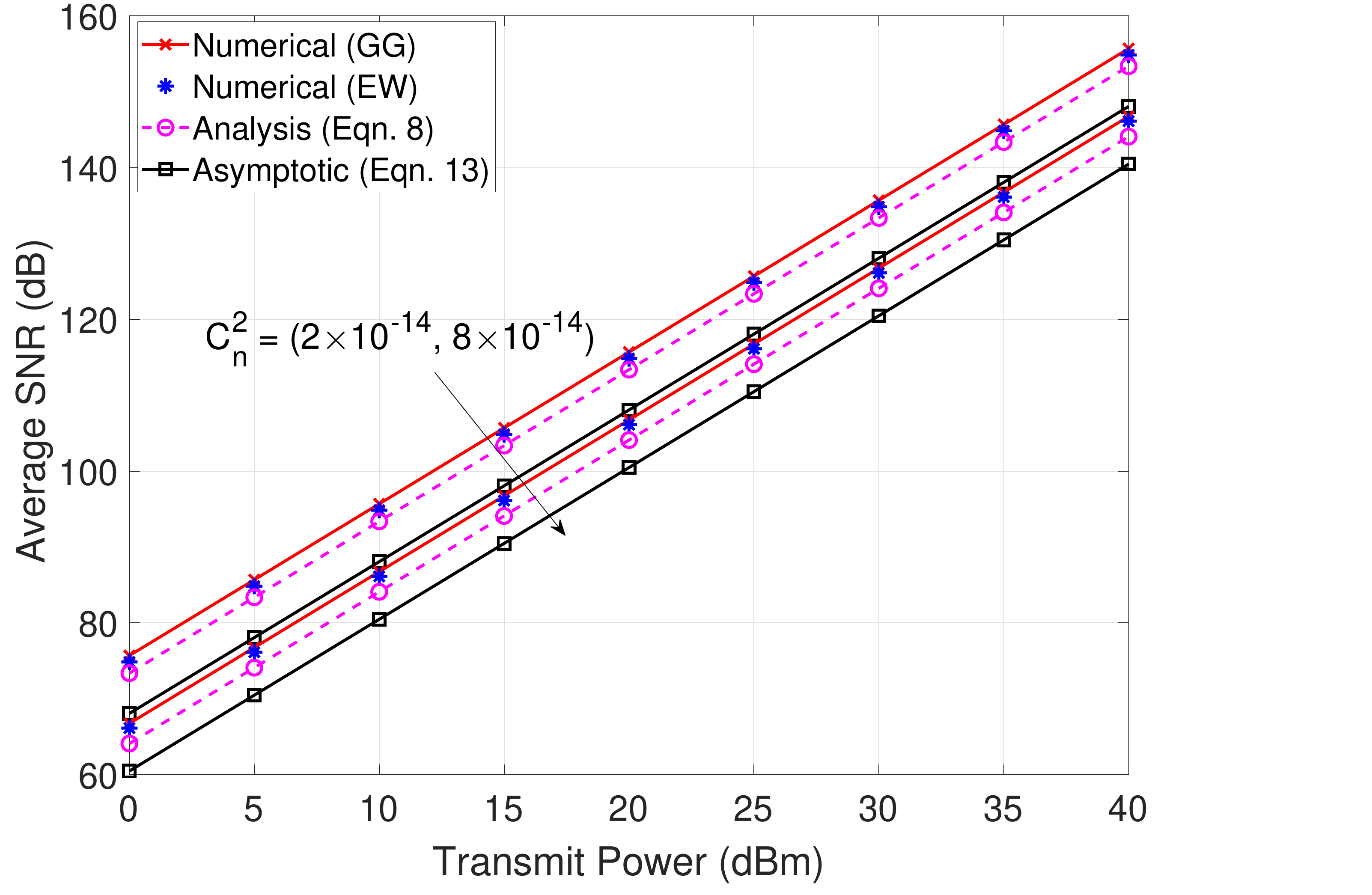}}\hspace{-2em}%
	\subfigure[Ergodic rate at $C_n^2= 2 \times 10^{-14}$,  $d=2$ \mbox{km}.]{\includegraphics[scale=0.19]{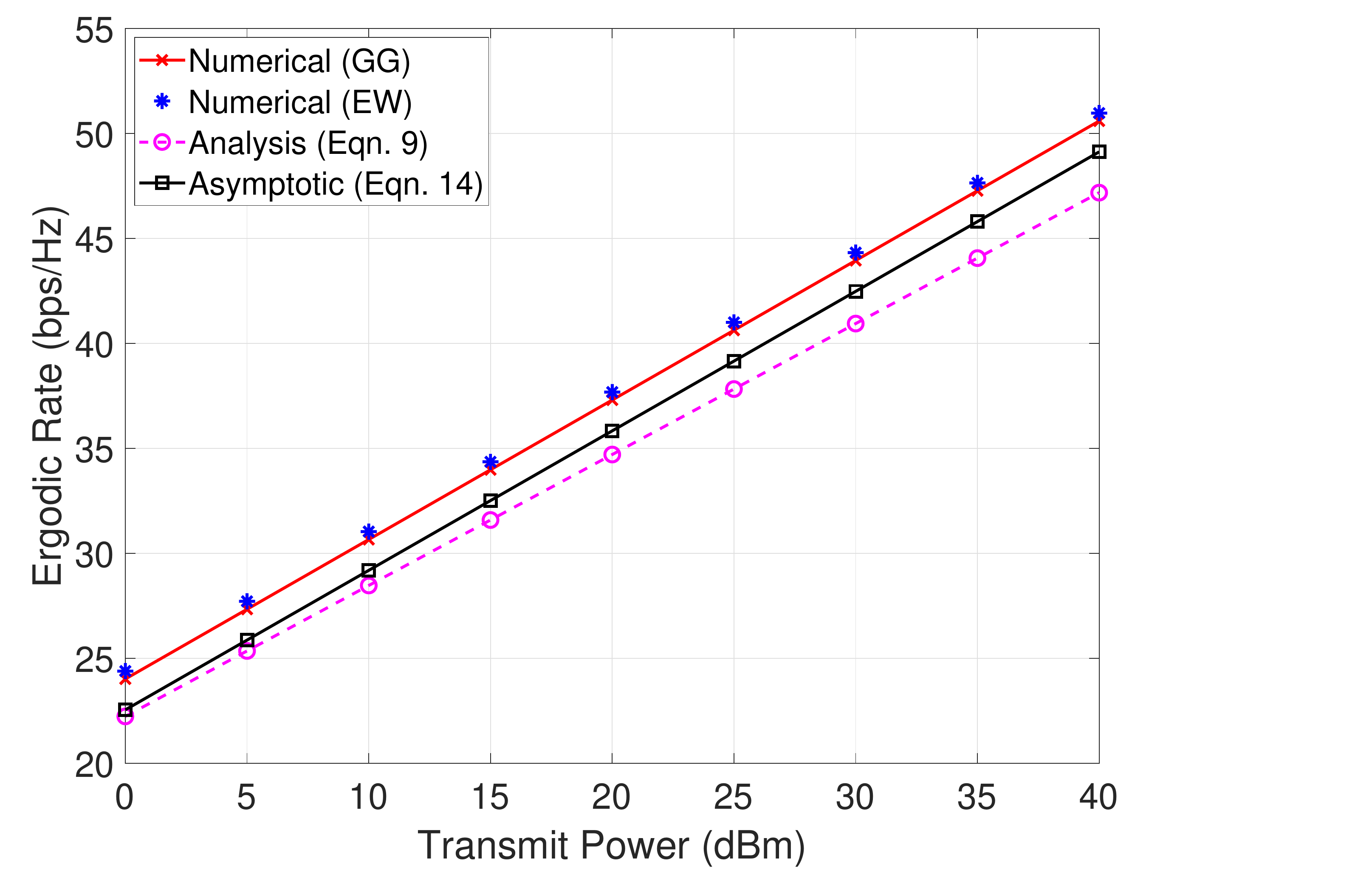}}
	\caption{Effect of atmospheric turbulence on  the average SNR and ergodic rate performance.}
	\label{fig:atm}
\end{figure*}

\begin{figure*}[t]
	\centering
	\subfigure[SNR at different link distances with $C_n^2= 8 \times 10^{-14}$.]{\includegraphics[width=\columnwidth]{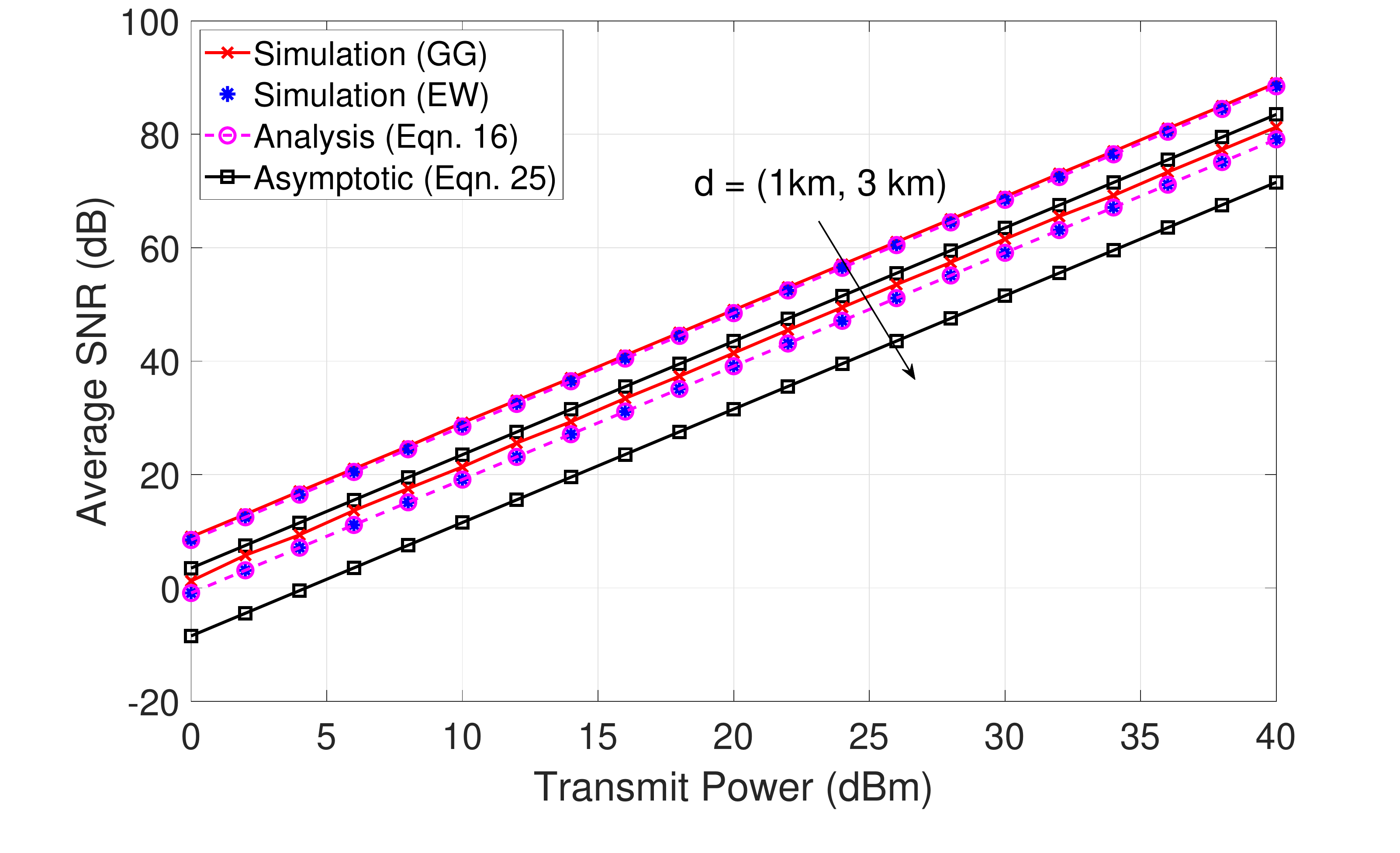}}\hfill
	\subfigure[SNR at different $C_n^2$ with $d=2$ \mbox{km}.]{\includegraphics[width=\columnwidth]{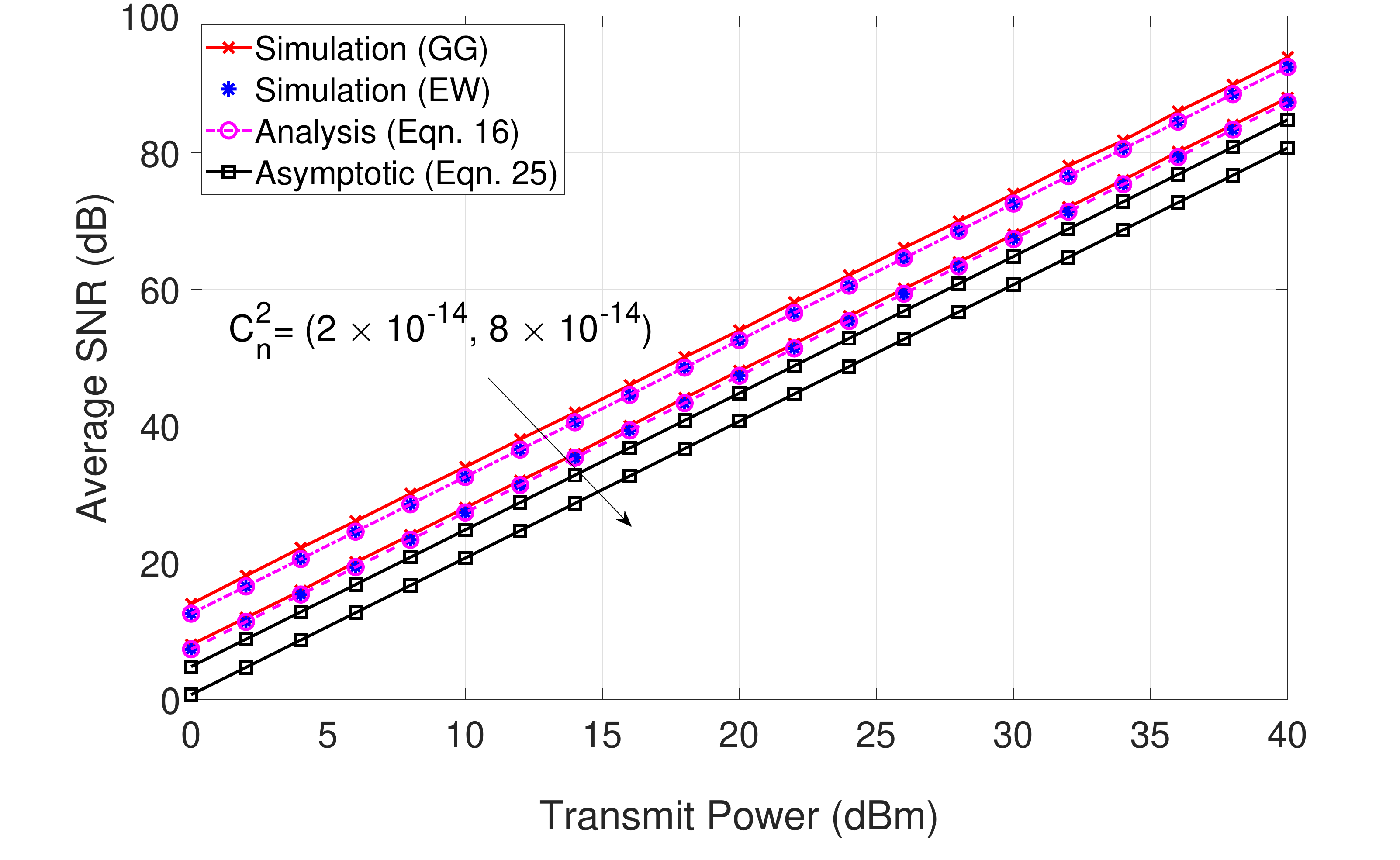}}
	\caption{Effect of atmospheric turbulence and pointing error on the average SNR performance. }
	\label{fig:snr_pointing}
\end{figure*}

	\subsection{Atmospheric Turbulence Channel with Pointing Error}

The PDF of  SNR under the combined effect of pointing error and atmospheric turbulence is given in  \cite{Sharma2015}:
\begin{align}
\begin{aligned}f_{\gamma}(\gamma)=B_{1} \sum_{j=0}^{\infty} \Psi(j) \gamma^{\frac{\rho^{2}}{2}-1} \Gamma\left[\tau, B_{2}(j) \gamma^{\frac{\beta}{2}}\right]\end{aligned}
\label{eq:pdf_SNR_atm_point}
\end{align}
where
$\Psi(j)=\frac{(-1)^{j} \Gamma(\alpha)}{j ! \Gamma(\alpha-j)(1+j)^{1-\frac{\rho^{2}}{\beta}}}$ , $ B_{1}=\frac{\alpha \rho^{2}}{2\left(L \eta A_{0} \sqrt{\bar{\gamma}_{0}}\right)^{\rho^{2}}}$, $\tau=1-\frac{\rho^{2}}{\beta}$ and $B_{2}(j)=\frac{1+j}{\left(L \eta A_{0} \sqrt{\gamma_{0}}\right)^{\beta}}$.  Here, {\footnotesize{$\Gamma(x)=\int\limits_{0}^{\infty}t^{x-1} e^{-t}dt$}} \normalsize is the Gamma function, and {\footnotesize{ $\Gamma(a,t) =\int_{t}^{\infty}s^{a-1}e^{-s}ds$}} \normalsize is the incomplete Gamma function. We also define $\psi(z)= \frac{d}{dz} \log (\Gamma(z))$ as the digamma function.
	\begin{my_lemma}
		\label{snr_atm_pe_exact}
		If $\rho$ and $A_0$ are the parameters of the pointing error, $\alpha$, $\beta$, and $\eta$ are the parameters of exponentiated Weibull turbulence channel and $L$ is the path loss of the OWC link, then   expressions for the average SNR and ergodic rate are given as 
		\begin{eqnarray}
		\begin{aligned}
		\overline{\gamma} = B_{1} \sum_{j=0}^{\infty}{\Psi(j) {\frac{2}{2+\rho^2}} {B_{2}(j)^{\frac{-2-\rho^2}{\beta}}}\Gamma\Big(\tau+ \frac{2 + \rho^2}{\beta}\Big)}                            
		\end{aligned}  
		\label{eq:SNR_atm_point_exact}
		\end{eqnarray}

			\begin{equation}
		\begin{aligned}	
		\overline{C}\geq &\frac{(-4) B_{1}}{(\log{2}) \beta \rho^{2}} \sum_{j=0}^{\infty}\Psi(j) B_{2}(j)^{\frac{- \rho^{2}}{\beta}} \Gamma(\tau + \rho^2/ \beta) \\
		& \Big( \beta + \rho^2 \log{B_{2}(j)} - \rho^2 \psi^{}{(\tau + \rho^2/\beta)}\Big)
		\end{aligned}  
		\label{eq:erg_cap_exact}
		\end{equation}
		
		\end{my_lemma}

	\begin{IEEEproof}	
	Using \eqref{eq:pdf_SNR_atm_point} in \eqref{siso_snr}, we get
		\begin{align}
	\overline{\gamma}= \int_0^{\infty } B_{1} \sum_{j=0}^{\infty} \Psi(j) \gamma^{\frac{\rho^{2}}{2}} \Gamma\left[\tau, B_{2}(j) \gamma^{\frac{\beta}{2}}\right]\, d\gamma
	\label{eq:pdf_SNR_atm_point_exp}
	\end{align}
	
To solve the above integral, we use  the following identity:	
	\begin{eqnarray}
		\int_{0}^{\infty} t^{a-1} \Gamma(b, t) \mathrm{d} t=\frac{\Gamma(a+b)}{a}, 
	 a>0, a+b>0
		\label{eq:gamma_inc_identity}
	\end{eqnarray} 
To do so, we substitute  $ t =\gamma^{\frac{\beta}{2}} $ in	\eqref{eq:pdf_SNR_atm_point_exp}, and apply the identity (\ref{eq:gamma_inc_identity}) in each term of the summation in  \eqref{eq:pdf_SNR_atm_point_exp} to get \eqref{eq:SNR_atm_point_exact}.	To derive \eqref{eq:erg_cap_exact}, we use \eqref{eq:pdf_SNR_atm_point} in  \eqref{siso_rate} and the inequality $\log_2(1 + \gamma) \geq \log_2(\gamma)$:
		\begin{eqnarray}
	\overline{C}= \int_0^{\infty } {B_{1}} \sum_{j=0}^{\infty} \Psi(j) \gamma^{\frac{\rho^{2}}{2} -1 } \Gamma\left[\tau, B_{2}(j) \gamma^{\frac{\beta}{2}}\right]
	\log_2(\gamma)	
	\, d\gamma
	\label{eq:pdf_erg_cap_atm_point_exp}
	\end{eqnarray}
	
	To solve the above integral, we use  the following identity:	
	\begin{eqnarray}
	\int_{0}^{\infty} t^{a-1} \Gamma(b, t) \log(t) \mathrm{d} t=&\frac{\Gamma(a+b) (-1 + a \psi ^{(0)}(a+b))}{a^2} 
	\label{eq:erg_cap_inc_ident}
	\end{eqnarray} 
Again, we substitute $ t =\gamma^{\frac{\beta}{2}} $ in 	\eqref{eq:pdf_erg_cap_atm_point_exp}, and apply the identity \eqref{eq:erg_cap_inc_ident} for each term of the summation in \eqref{eq:pdf_erg_cap_atm_point_exp}  to get \eqref{eq:erg_cap_exact}.
	
	\end{IEEEproof} 
	
The expressions \eqref{eq:SNR_atm_point_exact} and 	\eqref{eq:erg_cap_exact} are in closed form but they require infinite summation to get the exact results. Although the summation converges fast and only a few terms  are required to achieve a near-exact value, a different approach is required to derive a simpler bound. 	In order to do so, we consider the asymptotic PDF of the combined effect of atmospheric path loss, atmospheric turbulence, and pointing errors \cite{Boluda-Ruiz2017}:
	\begin{eqnarray}
	f_{h}(h)= \frac{\alpha \beta M_{r^2} \Big( \frac{2 \alpha \beta}{\omega_{z_{eq}}^2} \Big)  }{(L \eta A_{0}) ^ {\alpha \beta} } h^{\alpha \beta -1}, 0\leq h\leq D
		\label{eq:pdf_h_atm_point_asymp}
		\end{eqnarray}
	where $ D= \frac{L\eta A_0}{\big(M_{r^2}\big(\frac{2\alpha\beta}{\omega^{2}_{z_{eq}}}\big)\big)^{\frac{1}{\alpha \beta}}}$ and $M_{r^2}$ is the moment generating function  corresponding to the squared Beckmann distribution given as
	
	\begin{equation}
	\begin{aligned}
	M_{r^{2}}(t)=\frac{\exp \left(\frac{\mu_{x}^{2} t}{1-2 t \sigma_{x}^{2}}+\frac{\mu_{y}^{2} t}{1-2 t \sigma_{y}^{2}}\right)}{\sqrt{\left(1-2 t \sigma_{x}^{2}\right)\left(1-2 t \sigma_{y}^{2}\right)}}
	\end{aligned}
	\label{eq:M_r2}
	\end{equation}
	
	Here,  $\sigma_{x}$ and $\sigma_{y}$ represent different jitters for the horizontal displacement $x$ and the elevation $y$, and $\mu_{x}$ and $\mu_{y}$ represent different boresight errors in each axis of the receiver plane i.e.,	$x \sim N(\mu_{x}$, $\sigma_{x}$) and $y \sim N(\mu_{y}$ , $\sigma_{y}$).
	
Substituting $h =\sqrt{\frac{\gamma}{\gamma_0}}$ in \eqref{eq:pdf_h_atm_point_asymp}, we get an asymptotic  PDF of the SNR for an OWC system:
	\begin{eqnarray}
	f_{\gamma}(\gamma)= \frac{\alpha \beta}{2\sqrt{ \gamma \gamma_{0}} D^ {\alpha \beta} } \left(\sqrt{\frac{\gamma}{\gamma_{0}}}\right) ^{\alpha \beta -1},  0\leq\gamma\leq D^2\gamma_0
		\label{eq:pdf_SNR_atm_point_asymp}
	\end{eqnarray}

	\begin{my_proposition}
		\label{snr_atm_pe_approx}
			If $\rho$ and $A_0$ are the parameters of the pointing error, $\alpha$, $\beta$, and $\eta$ are the parameters of exponentiated Weibull turbulence channel and $L$ is the path loss of the OWC link, then asymptotic   expressions for the average SNR and ergodic rate are given as 
		\begin{eqnarray}
			\label{eq:avgSNR_asymp}
				\overline{\gamma} =& \frac{\gamma_{0} \alpha \beta}{2 + \alpha \beta} \Big(\frac{ M_{r^2} \big( \frac{2 \alpha \beta}{\omega_{z_{eq}}^2} \big)  }{(L \eta A_{0}) ^ {\alpha \beta} }\Big)^{1 -\frac{3}{\alpha \beta}} \;  \\                            
					\overline{C} =& 2 \gamma_{0} \frac{\alpha \beta M_{r^2} \big( \frac{2 \alpha \beta}{\omega_{z_{eq}}^2} \big)  }{(L \eta A_{0}) ^ {\alpha \beta} }  \left(\frac{\zeta}{\gamma_{0}}\right)^{\frac{\alpha \beta + 1}{2}} \Big(\frac{\alpha \beta \log{\zeta} -2}{\alpha \beta^2 \sqrt{\gamma_o \zeta } \log{4}}\Big)                       
			\label{eq:approx_erg_cap_asymp}
		\end{eqnarray}
		\end{my_proposition}
		\begin{IEEEproof}
It is straightforward 	to prove by  substituting 	\eqref{eq:pdf_SNR_atm_point_asymp} in 	\eqref{siso_snr} and 	\eqref{siso_rate}.	
	\end{IEEEproof}

 \section{Numerical and Simulation Analysis}
   
 This section demonstrates the average SNR and ergodic rate  performance of OWC systems using computer simulations. We also compare the performance obtained by the EW turbulence model with that of the GG model using numerical analysis as well as Monte Carlo (MC) simulations averaged by $10^6$ channel realizations.  We consider a wavelength of $1550$ \mbox{nm} (for path loss computation), detector responsivity $R=0.41$, and additive noise variance $10^{-14}$. The pointing error parameters are: receiver aperture diameter $2a=10$ \mbox{cm}, beam width $w_z= 2.5$ \mbox{m}, the maximum jitter $\sigma_x=\sigma_y= 35$ \mbox{cm}, and the maximum boresight $\mu_x=\mu_y=20$ \mbox{m}.  To analyze the effect of atmospheric turbulence, we consider the refractive index parameter $C_{n}^2=2\times 10^{-14}$ $m^{-2/3}$  with haze visibility of $V=4$ \mbox{km} for medium turbulence and $C_n^2= 8\times 10^{-14}$ $m^{-2/3}$ with  clear visibility of $V=16$ \mbox{km} for strong turbulence.

 \begin{figure*}[t]
 	\centering
 	\subfigure[Ergodic capacity at different link distances  with $C_n^2= 8 \times 10^{-14}$.]{\label{fig:a}\includegraphics[width=\columnwidth]{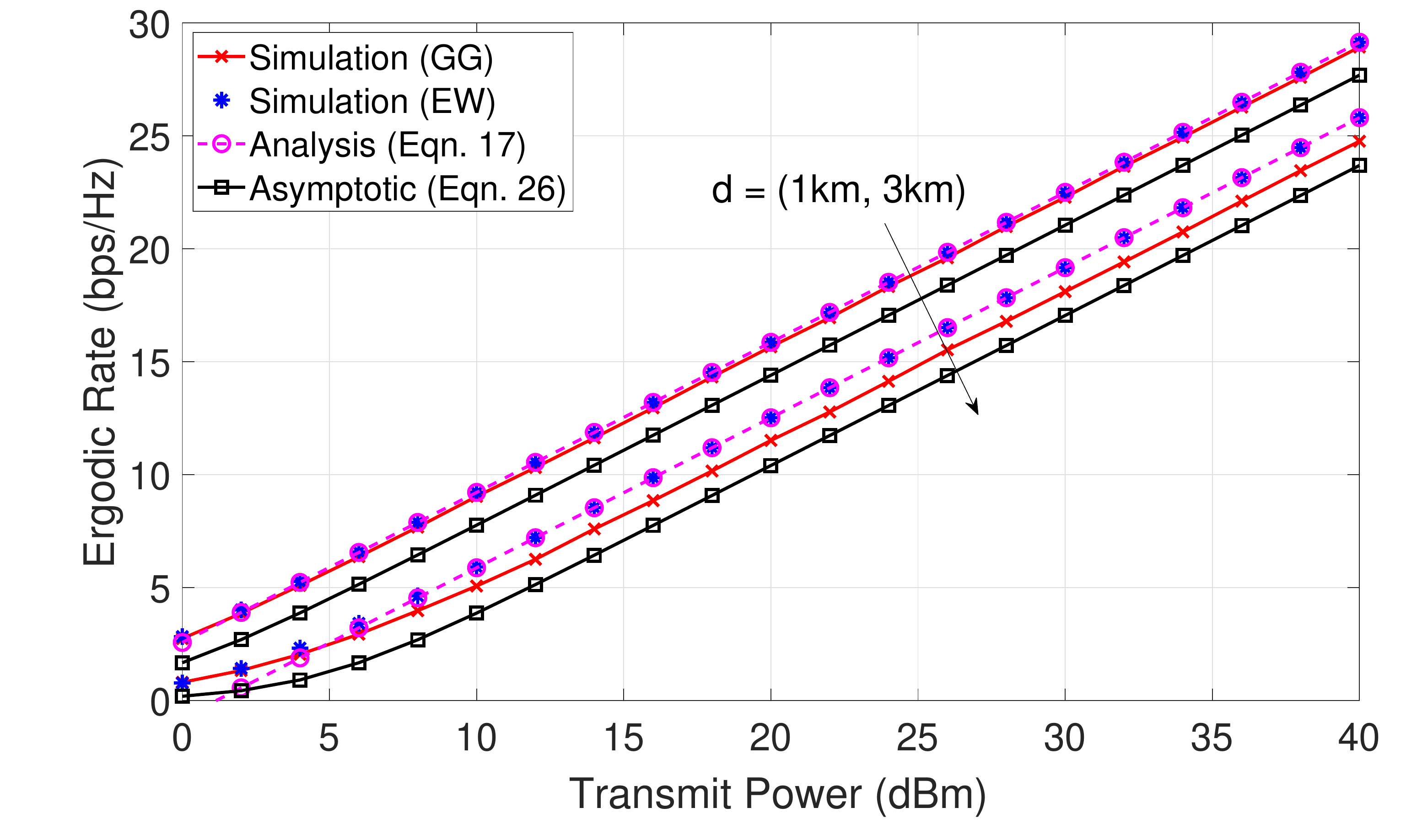}}
 	\subfigure[Ergodic capacity at different $C_n^2$  with $d=2$ \mbox{km}.]{\label{fig:b}\includegraphics[width=\columnwidth]{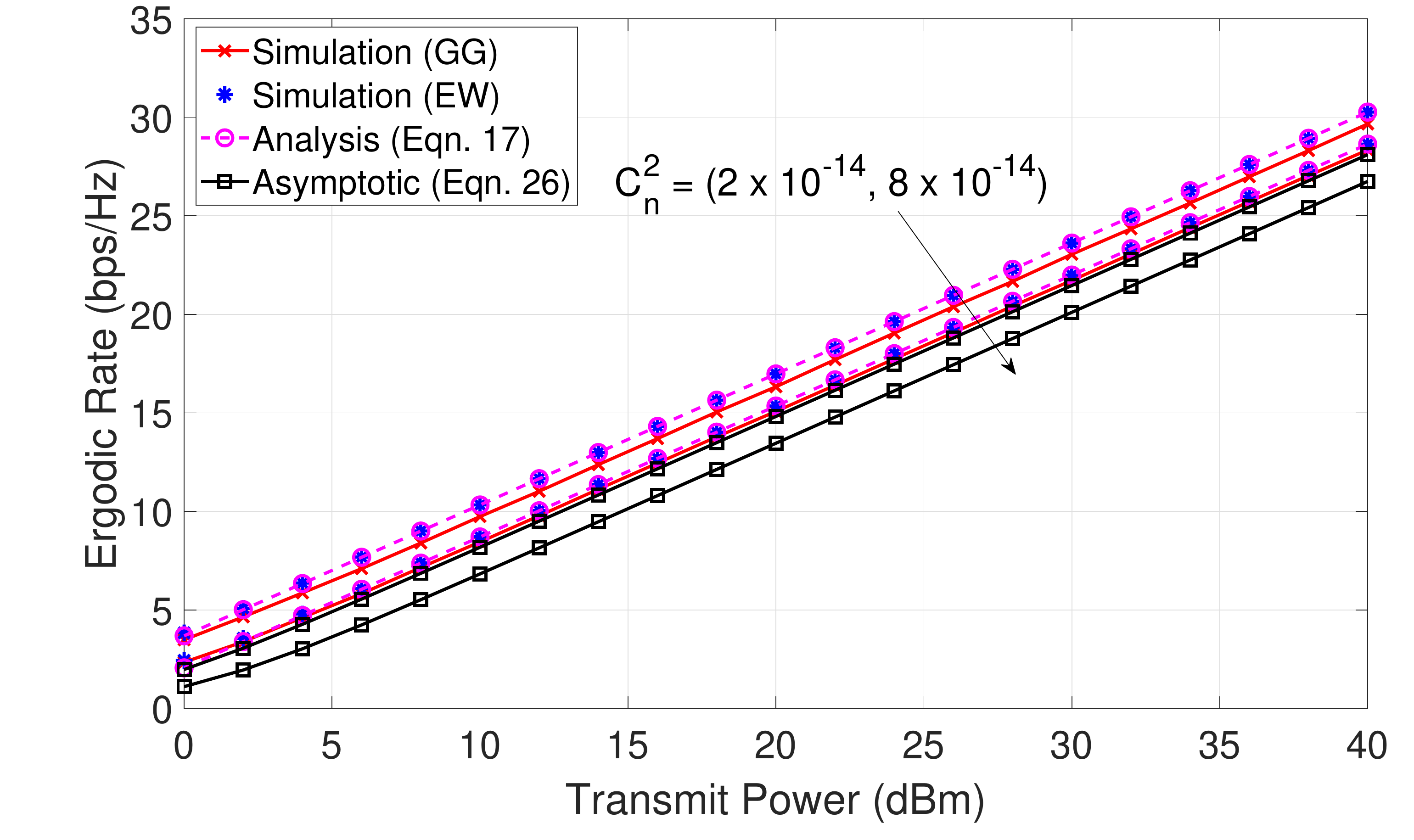}}
 	\caption{Effect of atmospheric turbulence and pointing error on the ergodic capacity  performance. }
 	\label{fig:cap_pointing}
 \end{figure*}

First, we demonstrate the effect of  turbulence channel on the average SNR and ergodic rate performance, as shown in Fig.~\ref{fig:atm}.
It can be seen from Fig.~\ref{fig:atm}a that the  average SNR decreases with an increase in  the link length, as expected. Moreover, stronger turbulence means higher visibility, and thus larger path loss, resulting a decrease in the average SNR  with an increase in the parameter $C_n^2$ (see Fig. \ref{fig:atm}b ). Further, the approximate and asymptotic analysis of the average SNR and ergodic rate (see Fig. \ref{fig:atm}c) is close to the exact results. It can also be seen that asymptotic analysis is closer to the approximate expression derived for the ergodic rate than the average SNR.

Next, we demonstrate the combined effect of atmospheric turbulence and pointing error on the average SNR. Compared with  Fig.~\ref{fig:atm}, the pointing error significantly degrades the average SNR performance, as shown in Fig.~\ref{fig:snr_pointing}. Considering the transmitted power of $22$ dBm, the pointing error reduces the average SNR by more than $60$ \mbox{dB} for a link distance of  $3$  \mbox{km}. 
It can be seen that our analysis in  \eqref{eq:SNR_atm_point_exact}) excellently matches  with the EW simulation results, as shown in Fig.~\ref{fig:snr_pointing}.  Furthermore, the derived asymptotic average SNR analysis  is closer to the exact results at a lower distance, as expected.

Finally,  Fig.~\ref{fig:cap_pointing} shows the ergodic capacity as a function of transmitted power for different values of link distances $d$ and refractive index parameter $C_{n}^2$.  The ergodic capacity of the system shows the similar trend as observed by the average SNR performance. The numerical evaluation of the derived expression matches closely to the  MC simulations except at a very low transmit power due to the use of inequality $\log_2(1+\gamma)\geq \log_2 (\gamma)$.

 In all the plots, it can be seen that the ergodic rate and average SNR  performance using EW turbulence model closely matches with the GG channel model advocating the EW fading model for performance analysis.

\section{Conclusions}

We have derived  simplified analytical expressions on the average SNR and ergodic capacity  performance of a single link OWC system by considering the exponentiated Weibull model for the atmospheric turbulence and Gaussian distribution model for misalignment errors. We have also presented asymptotic analysis to analyze the performance at higher SNR. Simulation and numerical analysis demonstrate the effect of atmospheric turbulence, pointing error, and visibility range on the performance of OWC system and verify the tightness of the derived expressions. The exponentiated Weibull fading is shown to be a potential  model for tractable performance evaluation since its performance excellently matches with that of the Gamma-Gamma model.
\section*{Acknowledgment}
This work is supported in part by the Science and Engineering Research Board (SERB), Govt. of India under Start-up Research Grant SRG/2019/002345.
\bibliographystyle{ieeetran}
\bibliography{ref_list_kartik}

\end{document}